# Building an Effective Data Warehousing for Financial Sector

**José Ferreira[1], Fernando Almeida[2], José Monteiro[1,*]**

[1]Higher Polytechnic Institute of Gaya, V.N.Gaia, Portugal
[2]Faculty of Engineering of Oporto University, INESC TEC, Porto, Portugal
*Corresponding author: almd@fe.up.pt

**Abstract**  This article presents the implementation process of a Data Warehouse and a multidimensional analysis of business data for a holding company in the financial sector. The goal is to create a business intelligence system that, in a simple, quick but also versatile way, allows the access to updated, aggregated, real and/or projected information, regarding bank account balances. The established system extracts and processes the operational database information which supports cash management information by using Integration Services and Analysis Services tools from Microsoft SQL Server. The end-user interface is a pivot table, properly arranged to explore the information available by the produced cube. The results have shown that the adoption of online analytical processing cubes offers better performance and provides a more automated and robust process to analyze current and provisional aggregated financial data balances compared to the current process based on static reports built from transactional databases.

**Cite This Article:** José Ferreira, Fernando Almeida, and José Monteiro, "Building an Effective Data Warehousing for Financial Sector." *Automatic Control and Information Sciences*, vol. 3, no. 1 (2017): 16-25. doi: 10.12691/acis-3-1-4.

## 1. Introduction

Although operational data is a key asset for an organization, such data is generally decentralized, inconsistent and not exploited to its full potential. In the context of this initiative we propose the development of a Business Intelligence (BI) project, implemented by the construction of a data warehouse (DW) and its Online Analytical Processing (OLAP) cube, for a holding company in the financial sector. These solutions, queries through a friendly interface, aggregated information regarding current and provisional financial data balances, by bank, company, country and account. In result, enable financial administration staff of the company to access to a homogenized and comprehensive view of the organization, supporting forecasting and decision-making processes at the enterprise level.

The origin of the information resides in a transactional database system that supports the accountability management department. Currently, the information is provided in the form of static reports, which are very inflexible and need a significant amount of work to be produced. Additionally, the data exploration features offered by the transactional application are very limited and time consuming, making a rapid response to new requests impossible.

Thus, we developed a DW, which was built using Microsoft technology already existing in the company, namely the Microsoft SQL Server (Integration Services and Analysis Services). In the end, the DW generates an OLAP cube information that can be exploited from Microsoft Excel.

The rest of the paper is organized as follows: First, we perform a revision of literature in the field of data warehousing by looking for three perspectives: data capture, data storage, and data access and analysis. After that, we briefly analyze the main DW initiatives and models for financial analysis. Then, we present the approach, methodology and phases adopted for the implementation of the project. Furthermore, we present the main results of the technical implementation of the project. Finally, we discuss the impact of these results in the activities of the holding company, and we draw the conclusions of our work.

## 2. Literature Review

The origin of the concept of data warehousing can be traced back to the early 1980s, when relational database management systems emerged as commercial products. The foundation of the relational model with its simplicity, together with the query capabilities provided by the Structured Query Language (SQL) language, supported the growing interest in business intelligence and decision support systems [1].

Since its inception, DW has been introduced in many industries, from manufacturing and production areas (order management and customer support), telecommunications and logistics, to health and financial services (analysis risk, fraud detection) [2].

As a consequence of the increasing changes in the current competitive world, the organizations need to perform sophisticated data analysis that supports your processes decision. Traditional databases typically associated with the operating systems do not meet the requirements for analysis of the information because they are targeted to support the daily basic operations. Therefore DW systems appear to be more suitable for the growing demands of decision-makers [3,4].

As referred by [5], a DW can be seen as an informational environment that: (i) provides an integrated and total view of the enterprise; (ii) makes the enterprise's current and historical information easily available for strategic decision making; (iii) makes decision-support transactions possible without hindering operational systems; (iv) renders the organization's information consistent; (v) presents a flexible and interactive source of strategic information.

A DW is not a single software or hardware product that can be purchased to provide strategic information [5]. It is, rather, a computing environment where users can find strategic information, and where they are put directly in touch with the data they need in order to make better decisions. For that reason, DW must be seen as a user-centric environment.

The building steps of a DW can be decomposed into three parts: (i) Data capture/acquisition, (ii) Data storage and (iii) Data access & analysis.

## 2.1. Data Capture/Acquisition

The acquisition component is the back end of the data warehousing system and consists of systems that have interface with the operational systems to load data into the DW [6]. Data is first entered or treated by a daily business process that is based on Online Transaction Processing (OLTP) environment and stored in operational database, which may consist of common relational databases such as Oracle, SQL Server, MySQL, DB2, etc. Before data is loaded from the operational database and external sources into the DW, it needs to be processed through three main functions: extraction, transformation and loading.

In the first phase of extraction, data are extracted from the available internal and external sources. A logical distinction can be made between the initial extraction, where the available data relative to all past periods are fed into the empty DW, and the subsequent incremental extractions that update the DW using new data that become available over time [7]. The selection of data to be imported is based upon the DW design, which in turn depends on the information needed for business intelligence analysis and decision support systems operating in a specific application domain.

The next step is the transformation phase that intends to improve the quality of the data extracted from the different sources, through the correction of inconsistencies, inaccuracies and missing values. Some of the major shortcomings that are removed during the data cleaning stage include inconsistencies between values recorded in different attributes that have the same meaning, data duplications, missing data and existence of inadmissible values [7]. Other common operations inside the transformation phase includes the conversion of character sets (e.g., from EBCDIC to Unicode), normalization/denormalization of the data into the desired dimensional DW schema, the assignment of surrogate keys to the data and maintenance of slowly changing dimensions by detecting changes and taking the actions necessary to contain with these [8].

In the loading phase the cleaned data is loaded into the tables of DW. Typically, the traditional DW architecture model assumes that new data loading occurs only at certain times, when the warehouse is taken offline, and the data is integrated during a more or less lengthy time interval [9].

## 2.2. Data Storage

The design of DWs is based on a multidimensional paradigm for data representation that provides at least two major advantages: on the functional side, it can guarantee fast response times even to complex queries, while on the logical side the dimensions naturally match the criteria followed by knowledge workers to perform their analyses [10,11]. There are two types of data tables in a multidimensional representation: dimension tables and fact tables. In general, dimensions are associated with the entities around which the process of an organization revolves. Dimension tables then correspond to primary entities contained in the DW, and in most cases, they are directly derived from master tables stored in OLTP systems. On the other hand, fact tables usually refer to a transaction and contain two types of data: links to dimension tables; and numerical values of the attributes.

A DW includes typically several fact tables, interconnected with dimension tables, linked, in their turn, with other dimensions. A fact table connects with n dimension tables and may be represented by an n-dimensional data cube where each axis corresponds to a dimension. Multidimensional cubes are a natural extension of the popular two-dimensional spreadsheets, which can be interpreted as two-dimensional cubes [12].

In order to standardize data analysis and enable simplified usage patterns DWs are commonly organized as problem-driven, small units, called "data marts", each data mart is dedicated to the study of a specific problem [13]. A data mart can be considered as a functional or departmental DW of a smaller size and more specific type than the overall company DW. As a consequence, a data mart contains a subset of the data stored in the company DW, which are usually integrated with other data that the company department responsible for the data mart own and deems of interest.

In order to understand and locate data in the DW users need information about the data warehousing system and its content. This information is known as metadata. The metadata indicates for each attribute of a DW the original source of the data, their meaning and the transformations to which they have been subjected. The metadata should also include business definitions, data quality alerts, organizational changes, business rules and assumptions [14].

## 2.3. Data Access & Analysis

The access component of a DW project is referred as the front end. It consists of access tools and techniques that provide a business user with direct, interactive, or batch access of data, while hiding the technical complexity of data retrieval. The interface provides an intuitive, business-like presentation of information, friendly enough for a non-technical person. A variety of tools can be typically used, such as data analytical tools, data mining, machine learning, etc.

In order to facilitate complex analysis and visualization, the data in warehouse is typically modelled multidimensionally. The best known knowledge discovery techniques are Online Analytical Processing (OLAP) and data mining (DM) techniques [15].

OLAP provides users with the means to explore and analyse large amounts of data, involving complex computations, their relationships, and visually present results in different perspectives. Typical OLAP operations include rollup (increasing the level of aggregation) and drill-down (decreasing the level of aggregation or increasing detail) along one or more dimension hierarchies, slice and dice (selection and projection), and pivot (re-orienting the multidimensional view of data) [2].

Data mining is the process of discovering insightful, interesting, and novel patterns, as well as descriptive, understandable, and predictive models from large-scale data [16]. The relationships and summaries derived through a data mining exercise are often referred to as models or patterns. Examples include linear equations, rules, clusters, graphs, tree structures, and recurrent patterns in time series [17].

## 3. DW Initiatives and Models for Financial Analysis

A DW is set of historical data, sometimes obtained from different sources, and its main purpose is to support management decisions. Its implementation is a complex process, namely when one's problematic situation is different from other known implementations. Thus, it is important to take into consideration the information needs of decision makers (requirement-driven approaches), but also what data are available as a source of information to populate the DW (data-driven approaches). These two perspectives, with the necessary adaptations lead us to a hybrid approach [18]. This implies the needs of the management to rely on the capability of exploring available data, considering high volumes of data, such as in a big data scenarios or just looking to a way of organizing the information in a credible an consistent form [19].

Literature also refers several proposals on multidimensional modelling and data warehouse design [8,11,20]. However, it seems, there is no consensus on modelling and design method, but a set of complementary methods [21]. The Conceptual Data Model for Data Warehouse of Kamble, points to a uniform way of modelling multidimensional concepts, data warehouse design and aggregations [21]. The influence of this model on our project is at a level of research, which individually, focus on a small part of the problematic situation. In fact, each model referred by Kamble focus on a particular approach about managing data.

In the research on the business domain of the project we stated that finding studies of DW to solve the needs of bank entities seems to be common. However, studies of DW to link the bank accounts to the financial area of a financial company seem to be a new problematic situation. To turn around this adversity, we conduct our study by researching conceptual models that could inspire the design and the implementation of DWs in a broad sense.

The schema of a DW lies on two kinds of elements: (i) facts and (ii) dimensions [21]. Facts are related with the metrics of situations or events, and dimensions are used to analyse the results of the metrics, through the application of a set of arithmetic operations (counting, summation, average, ...).

Establishing an analogy with our case, the movements of bank accounts are the facts and, each fact is characterized by a monetary value and by the value of the situation (debt or credit). The monetary values can be aggregated by debt or credit. However, other dimensions can be obtained: balance by year, balance by month, debts higher than a certain value, credits from a specific entity, among other dimensions. So, the value of the movement is the connection between the types of the movement with the restriction that filters a set of movements. In the simplest way, this type of connection concerns the organization of facts with regard to dimensions [22]. The model of Schneider, has brought the unification of the notion of fact and the notion of dimension, which become a value to future implementations of DWs. Another valuable characteristic is the similarity among model representation and semantic web. Like the "Conceptual Data Model" of Kamble, it allows to solve adversities based on the reasoning of the semantic web [21]. Both are related with mathematical theoretic semantics grounded on standard ER semantics.

## 4. Approach

### 4.1. Methodology

Due the purposes of this project we characterize our methodology according two perspectives: (i) goal-oriented – the project intends to solve a problem related with integration of cash movements of business establishments; (ii) user-driven – the project has strong support of users in order to meet its requirements [23].

The methodology adopted in his work has four stages and seven phases (Table 1). Each stage includes one or more phases: (s1) definition; (s2) execution; (s3) exploration; (s4) tests.

**Table 1. Methodology stages and phases**

| Stages | Phases |
| --- | --- |
| s1 - Definition | Requirements |
|  | Architecture |
|  | Multidimensional model |
| s2- Execution | Integration services |
|  | Analysis services |
| s3- Exploration | Spreadsheet for inf. retrieved |
| s4 - Tests | Tests to evaluate results |

The definition stage (s1) was divided into three phases: (i) in the first phase was defined the core of the project and the requirements to a successful implementation (functional and non-functional requirements definitions); (ii) in the second phase was defined the architecture of the project. It was decided to build the DW based on Microsoft technology, namely Microsoft SQL Server 2008 and its Integration Services (SSIS) and Analysis Services (SSAS) components; (iii) in the third phase was defined the Multidimensional model.

In the execution stage (s2) were included two phases: (i) execution of the Integration Services phase; (ii) execution of the Analysis Services phase

During the exploration stage (s3) was designed a Spreadsheet for information retrieved and exploration.

Finally, in the tests stage (s4) were performed a set of tests in order to evaluate the results.

For better understanding of the procedures performed during the stages of the project, each phase will be described in the next sub-sections.

### 4.2. Functional Requirements

This project is organized in seven stages: Functional and Non-Functional Requirements definition, System Architecture definition, Multidimensional Model definition, Integration Services Phase execution, Analysis Services Phase execution, excel spreadsheet elaboration to explore information and, finally, a test phase.

The functional requirements specify what the system should do to meet the user's needs. This project intends to enable the consultation of real account balances (based on the bank account statements), but also the consultation of account balances that include estimated account movements (known as working balances). These balances can be aggregated by owner account, by bank, by currency, by company country and bank country. Additionally, the system should allow the setting of temporal criteria, which means, to specify the day, or a time scope, which can be weeks, months, quarters, semesters or years. It should be possible to aggregate the balances by sum or average.

### 4.3. Non-Functional Requirements

The non-functional requirements specify how the system should behave and it is typically responsible for the definition of the overall qualities or attributes of the application. It places restrictions on the product being developed, its process, and specifies external constraints that the product must meet.

The following non-functional requirements were established within the context of this application:

○ Availability – the application should be always available in order to allow that top management query the data and draw conclusions when appropriate. It is considered especially critical the availability on weekdays;

○ Performance – the application must offer short response times, i.e., the information must be available within few seconds since it is requested a spreadsheet update with data from the DW;

○ Usability – the organization of information in the decision-cube should be logical and intuitive in order to be simple, fast and easily interpreted and perceived by its users. It is intended that a user through a brief training of 5 minutes, should be able to use all the application functionality in less than 15 minutes;

○ Security – the access to information that the solution provides to the end user should only be allowed to grant rights, in particular, to the staff of Administration and Financial Department. The access must be read only.

The portability was not considered as a mandatory non-functional requirement for this project, because users will interact with the system using only desktop interfaces.

### 4.4. Architecture

The system consists in a data warehousing solution, built using Microsoft technology, specifically Microsoft SQL Server 2008 and its Integration Services (SSIS) and Analysis Services (SSAS) component. The information provided by the SSAS project, consists in an OLAP cube which can by access by the final user using a Microsoft Excel 2010 spreadsheet, properly connected to the SSAS. A visual representation of the architecture is depicted in Figure 1.

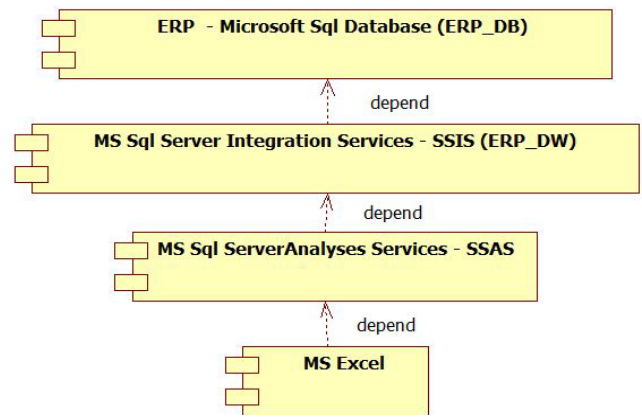

**Figure 1.** System architecture

### 4.5. Multidimensional Model

In this stage we firstly describe the data source. The following points address the definition of the facts table, level of granularity, dimensions e the relational schema.

1) Data Source

All the data that is gathered to integrate the DW is coming from a transactional database, in SQL format, that supports the software application for treasury management and treasury operations. The only exception is the time table as explained later.

2) Fact table

As Figure 2 shows, the fact table contains the value date and account attributes (that form the primary key). It also contains attributes for the foreign keys from each of the dimension tables shown below, as well as quantitative attributes like account balances in euro and original currency of the transaction, and the working account balances in euro and in original currency.

3) Granularity

The maximum detail level of the fact table records corresponds to daily balance of each account, which results from the aggregation of all movements occurred in each day, for each account.

4) Dimension Tables

The defined dimension tables are: Companies, Banks, Accounts, Currencies, Countries and Time,

5) Relational Schema

The adopted schema, represented in the class diagram in Figure 2, is mapped in a star schema where each dimension table is directly connected to the fact table.

## 4.6. Integration Services Phase

The SSIS project consists in the creation of a data source package that establishes the data flow tasks that extract, transform and load the data into the database (ETL). These tasks are aggregated in four sequence containers, as shown in the Figure 3.

A time table was previously created with one record for each day between 2009 and 2016. Each record contains the date's week, month, quarter, semester and year, and also indicates if that date is the last day of the period (week, month, quarter, semester and year). The time table is managed by the DW administrator. By the end of each year, this table should be added with a set of records that corresponds to one more year of records, on for each day. For example, in the end of 2016, the time table should be added with 365 records, one for each day of 2017.

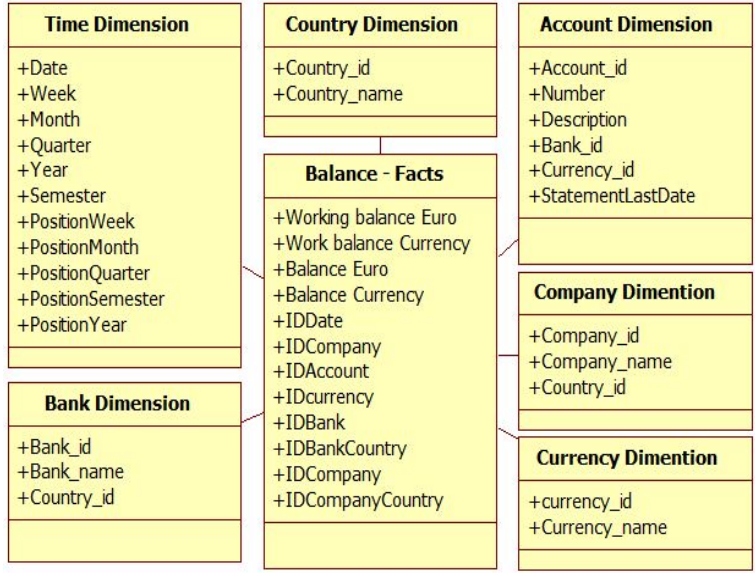

**Figure 2.** Class diagram of the DW

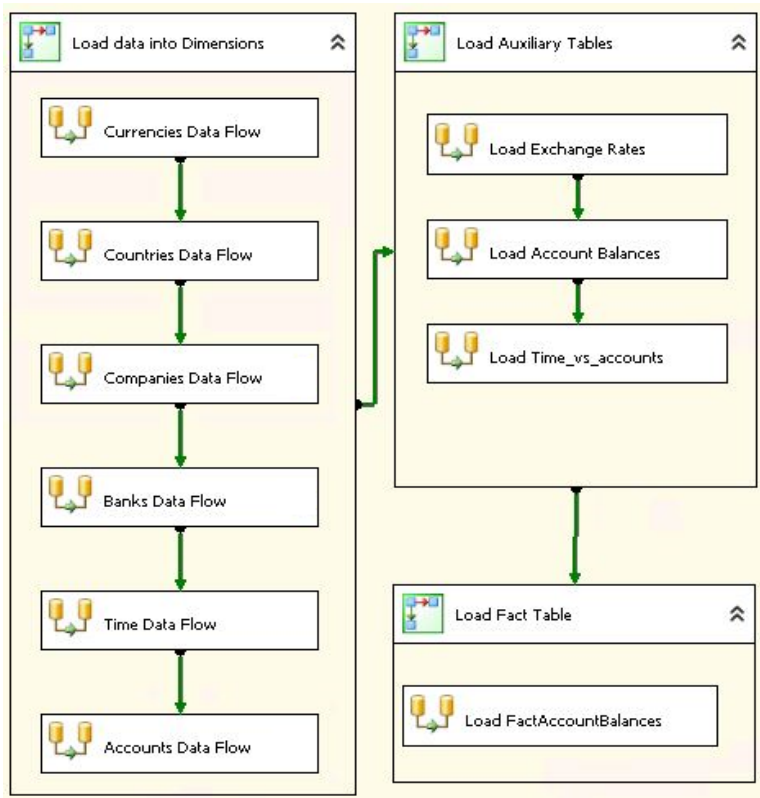

**Figure 3.** Data flow tasks of SSIS project

Figure 3 represents graphically the data flow tasks of the SSIS project.

1) Load data into Dimensions

Each sub-task includes the insertion in the dimension tables of the data coming from the treasury management software's database, using SQL syntax.

2) Load Auxiliary Tables

This phase consists in loading data into auxiliary tables that are not represented in the dimensional model, trough the following sub-tasks:

a) Load updated information related to the exchange rates;

b) The sub-task that follows, which is represented in detail in Figure 4, comprises charging to an auxiliary table called "Account Balances", the balances for each account on each day that occurred a movement. For this purpose, this process collects current movements (real), the opening balances (actual) and forecast movements. Furthermore, it converts the value-data in the correct format calling the tool "derived columns", identify the currency using a "lookup" tool associated to the currencies table, aggregate in sum the total balance (treasury) and from sheet movements (only actual movements) for each account and different value-data, thus resulting in the everyday balance;

c) It follows the loading of another auxiliary table called "Time vs Accounts" that receives the result of the joining between the account table with the time table. This table constitutes the support base structure to the calculation of daily sales because it contains a record for each account/day. Without this structure, DW would contain only balances on days that had movements, thereby making it difficult to query for a balance in day without movement and, on the other hand, the average balance calculation.

3) Load Fact Table

The last phase is the creation of fact table called "FactAccountBalances". This process merges the table "Time vs Accounts" with the "Account Balances", thus resulting in a record with the balance of each account in each day of the time table (and not just on the days when there was movements). The fact table is supplemented with complementary information regarding companies and banks by using the "lookup" tool associated with the account table.

Since the available exchange currency rate in the system, needed to calculate the account balance in Euros, is two days overdue, the account balance conversion to Euro should be updated every day, considering the most updated information about the exchange currency rate. So, although DW is characterized by data volatility, this is, once data is integrated it will no longer be modified, it was decided to update facts if the values in Euros are different, by assuming that the new value is more correct.

Therefore, this process verifies if each fact already exists in the fact table by comparing the two key columns: due date and account. If the fact does not exist, then it is loaded into the fact table. Otherwise, the process checks if the account balance in Euros and the working account balances in Euros are different from the values in the existing fact in the fact table. If so, then the fact is updated with the new values, otherwise the fact is not loaded since it already exists and the values are equal.

**4.7. Analysis Services Phase**

Based on compliance with the following steps:

1) Data Source Definition and Data Source View

It was created a view with the dimensions and fact tables from the resulting DW project SSIS. These tables are organized using the star schema already presented. The above mentioned auxiliary tables were not considered because they are not necessary.

2) Creation of Dimensions

From the Data Source View was created and characterized the dimensions that will be considered in the cube. One should mention the specific case of the time dimension whose type was defined as a time and not as regular as the others. This is particularly important because it enables the definition of measures based on time such as the average balance.

3) Creation of Cube

From the created dimensions and existing measures in the fact table (the various types of balances) was established an information cube that was complemented with an additional measure: average balance sheet in Euros. This measure belongs to average over time type.

Finally, an automatic task was created, on a daily bases, that updates the DW and its information cube.

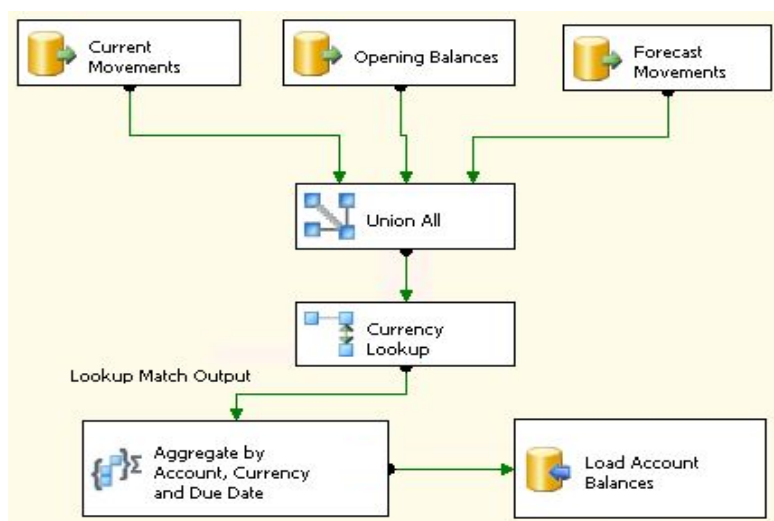

**Figure 4.** Data flow balances

### 4.7.1. Spreadsheet to Data Cubes Exploration

A spreadsheet in MS Excel 2010 properly connected to the source of the Analysis Services data type was created. In this sheet several pivot tables were created (PivotTable) in order to meet the functional requirements.

To check the accuracy of the information presented, unit tests were performed in order to compare against the results from the operational treasury management software.

## 5. Results

Table 2 and Table 3 have a similar structure and intend to present the results obtained for each use case. In this structure, the first section lines show the results from the modality I: "Use a PivotTable in an Excel spreadsheet, previously parameterized and whose source of information is the OLAP cube, where the "refresh" operation can be used to update the information". The second section lines show the results from the modality II: "Method used before the implementation of the OLAP cube, which involves extracting the daily balances from the cash management program, export the data to Excel and update a spreadsheet Excel previously prepared to present the required information.

For each modality we measure the arithmetic average and standard deviation of three measurements, with a millisecond precision level. A comma was used to represent a decimal point.

**Table 2. Estimation of account balances in next month and year**

| Get estimated account balances in next: | Month | | Yesr | |
|---|---|---|---|---|
| | Avg. | Std. | Avg. | Std. |
| 1. Through OLAP cube | | | | |
| Refresh * | 0,827 | 0,045 | 0,915 | 0,027 |
| 2. By the method previously used | | | | |
| Get daily balances from operational application * | 11,569 | 0,895 | 60,305 | 0,565 |
| Export data to Excel * | 11,645 | 0,620 | 15,338 | 0,393 |
| Total * | 23,215 | 1,515 | 75,643 | 0,958 |
| Time benefit differences | | | | |
| 1 day | -0,4 minutes | | -1,2 minutes | |
| 1 month = 22 working days | -8,2 minutes | | -27,4 minutes | |
| 1 year = 22 working days * 12 months | -1,6 hours | | -5,5 hours | |
| Hypothesis testing | | | | |
| difference in means (t student) | 25,590 | | 135,019 | |
| t-value (2-tailored 95%) | 2,776 | | 2,776 | |
| t-value (2-tailored 99%) | 4,604 | | 4,604 | |
| degrees of freedom | 4 | | 4 | |
| Number of considered forecasts | 337 | | 2553 | |

* in seconds with an accuracy of milliseconds.

**Table 3. Getting the average balance from all accounts for all months since last 3 years**

| Get estimated account balances in next: | Last year | | Last 3 years | |
|---|---|---|---|---|
| | Avg. | Std. | Avg. | Std. |
| 1. Through OLAP cube | | | | |
| Refresh * | 1,015 | 0,025 | 1,260 | 0,046 |
| 2. By the method previously used | | | | |
| Get daily balances from operational application * | 65,652 | 0,923 | 324,35 | 2,052 |
| Export data to Excel * | 16,357 | 0,774 | 34,57 | 0,947 |
| Total * | 82,009 | 1,698 | 358,92 | 2,999 |
| Time benefit differences | | | | |
| 1 day | -1,3 minutes | | -6,0 minutes | |
| 1 month = 22 working days | -29,7 minutes | | -131,1 minutes | |
| 1 year = 22 working days * 12 months | -5,9 hours | | -26,2 hours | |
| Hypothesis testing | | | | |
| difference in means (t student) | 82,632 | | 206,523 | |
| t-value (2-tailored 95%) | 2,776 | | 2,776 | |
| t-value (2-tailored 99%) | 4,604 | | 4,604 | |
| degrees of freedom | 4 | | 4 | |
| Number of considered forecasts | 27688 | | 67888 | |

* in seconds with an accuracy of milliseconds.

In the third section lines the difference in time spent in favor of modality I is displayed, which is obtained by subtracting the mean time spent in modality I, the average time spent in modality II, for each of the following three scenarios: a) per day; b) per month, considering 22 days per months (working days); c) per year, considering 264 days (22 days multiplied by 12 months). For each scenario we considered that the use case is performed once per day.

In the fourth section we perform a hypothesis testing regarding the difference of the two execution times means (Modality II - Modality I). For that, we adopt a t-student test, considering that the total number of samples is less than 30, and we register the t-value (2-tailored) for a confidence level of 95% and 99%.

Finally, it should be noted that we used a personal computer with Intel Core i3 2.3GHz processor with 4 GB RAM to perform those performance tests. The computer is equipped with Windows 7 32-bit and were installed the SQL Server database and all the programs mentioned in the system architecture section.

### 5.1. Use Case I - Estimation of Account Balances

Table 2 presents the measurements of time spent in the following use case: obtain estimated account balances for next month and next year, based on the known balances by the end of the day.

A comparative graphical analysis between the measurements of the time spent in the estimation of account balances in next month and next year through OLAP cubes is given in Figure 5.

### 5.2. Use Case II - Getting the average balance from all accounts for all months

Table 3 presents the measurements of time spent in the following use case: receive the average balance for each account and each month, based on the known balances by the end of the day.

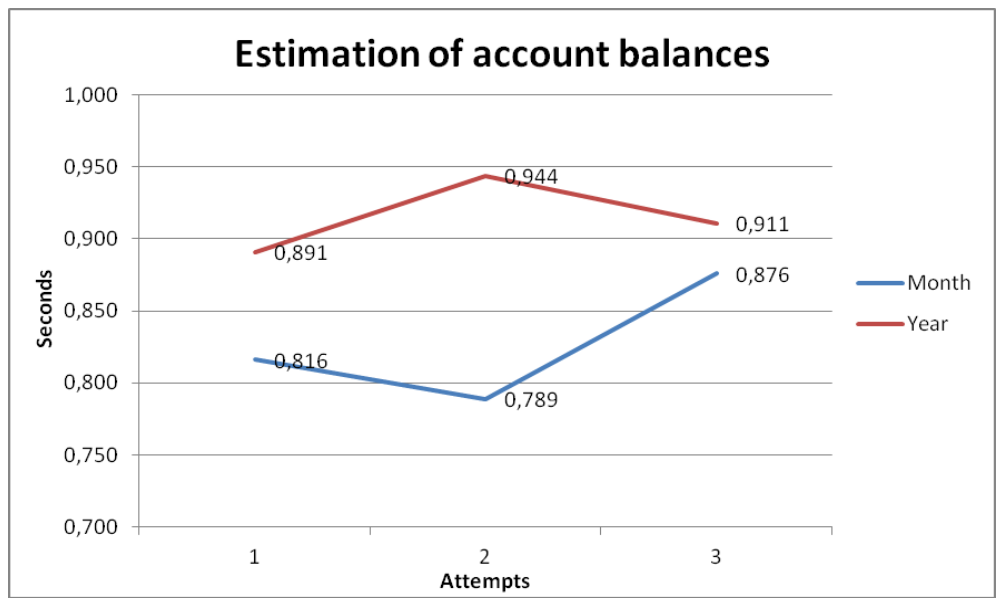

**Figure 5.** Estimation of account balances in next month and year

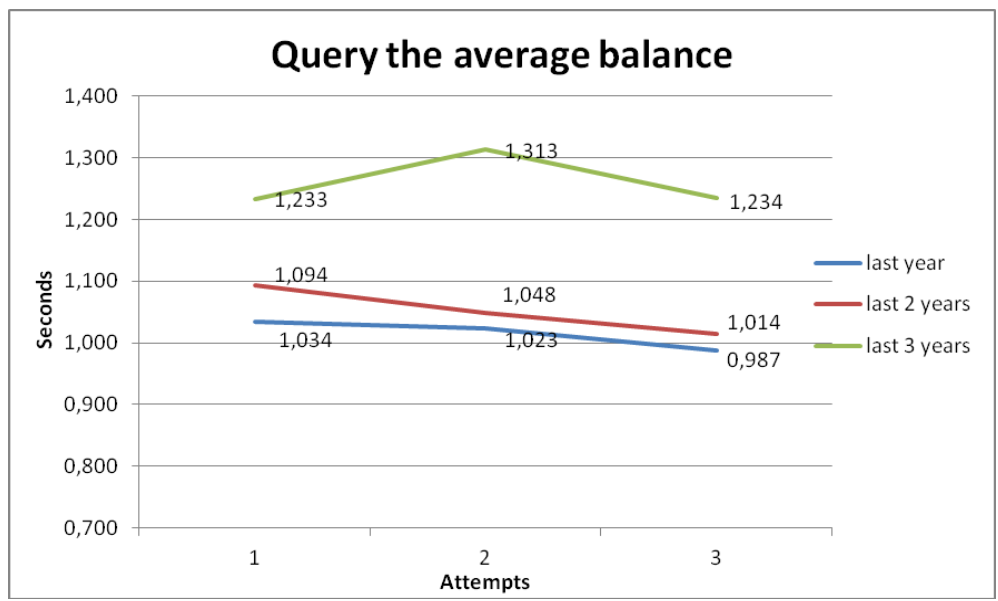

**Figure 6.** Query the average balance since last 3 years

A comparative graphical analysis between the measurements of the time spent getting the average balance from all accounts for all months through OLAP cubes is given in Figure 6.

## 6. Discussion

Looking at the results section, obtaining information in both use cases was shown to be faster through the use of an OLAP cube when compared with the method previously used by the holding company. Looking for the first use case, the hypothesis testing for the difference between the two means (OLAP cube vs. traditional method) is higher than the t-value (2-tailored) for a significant level of 0,05 and 0,01. Therefore, this difference can be considered extremely statistically significant, which means that for use case I the use of OLAP cube offers better performance. This conclusion is valid for both scenarios (estimation account balances in next month and next year). In the second use case, the hypothesis testing for the difference between the two means (OLAP cube vs. traditional method) is also higher than the t-value (2-tailored) for a significant level of 0,05 and 0,01. Consequently, this difference can be also considered extremely statistically significant, which means that the OLAP cube offers better performance. This conclusion is valid for the three considered scenarios, when we get the average balance from all accounts and for all months since last year, last 2 years and last 3 years.

We also estimated the time benefits differences by the adoption of OLAP cubes. In use case I, the saving time by using OLAP cubes is around 5,5 hours after one year of use, which is equivalent to almost one working day (considering 7,5H/day). In use case 2, the saving time by adopting OLAP cubes is 26,2 hours after one year of use, which is equivalent to more than three business working days. If we consider that these use cases have a daily recurrence, the sum of the time savings of these two cases by employing OLAP cubes is more than 4 working days per day. Additionally, the growth of the number of records involved in the analyzed data (number of years) was shown to have a greater impact in use case I than in use case II. In other words, the greater the amount of information to be examined, the greater is the advantage of adopting OLAP cubes, which also confirms the findings obtained by Chouhan [24].

Finally, the increase of the number of records involved in the analyzed data (number of years) can be observed as having a greater impact in embodiment 2 than in embodiment 1. Simply put, the greater the amount of information to be examined, the greater is the advantage mode 1.

## 7. Conclusion

The performed work demonstrated that the use of a DW approach is an important tool to support decision-making and for the strategic management level, both the tactical and even operational level. The concept of building a DW presented in the literature review section was performed in practice using Microsoft SQL Server BI tools. The use of OLAP cubes facilitates the retrieval and analysis of large amounts of data. The Excel displays summarized data in PivotTable reports and OLAP server performs calculations to summarize the data. The end user interface is simple to use and allows the user to create analyzes with great flexibility (by month, year, bank, business type, real or estimated balance, etc.). Additionally, the analysis of the process becomes more automated and robust to the introduction of errors in the analysis of the process by the user. These are in fact the great advantages of this study to overcome the difficulties that existed before, wherein the access of information and analysis were very costly in terms of time and offered very limited flexibility.

In the future it is planned the development of this working. Currently we have identified two needs: the identification of accounts with infringement balances (based on pre-defined parameters); analysis of aggregate movements by budget code.